\documentclass[
preprint,
eqsecnum,
aip,jmp,
amsfonts,amssymb,amsmath,
longbibliography
]{revtex4-1} 

\usepackage{hyperref}
\usepackage{bm}
\newcommand{\du}{\mathrm{d}}

\begin{document}
\title{Lorentz transformation of electromagnetic pulses derived from Hertz potentials}
\author{Petar Andreji\'{c}}
\affiliation{
  School of Chemical and Physical Sciences,
  Victoria University of Wellington,
  PO Box 600,
  Wellington,
  New Zealand
}
\date{\today}

\begin{abstract}
  Electric and magnetic Hertz potentials are a formalism for obtaining solutions of Maxwell's equations from solutions of the inhomogeneous wave equation, with polarisation and magnetisation as the sources. We provide an overview of their covariant transformation properties, and examine an application of Hertz potentials to the Lorentz transformation of localised pulses by Lekner, who obtained a result seeming to contradict Von Laue's theorem. We show that Lekner's result of total energy-momentum not transforming as a four-vector was due to an erroneous transformation of Hertz potentials, that did not take into account their bivector nature.
\end{abstract}

\maketitle

\section{Introduction}\label{sec:introduction}

In the covariant formulation of electromagnetism, the energy and momentum density do not constitute a four vector, and instead are components of the electromagnetic stress-energy tensor, $T^{ab}$ (Jackson\cite{jackson_classical_2007}, Chap.~12). Similarly, the angular momentum is not an anti-symmetric rank 2 tensor, but is instead given by the rank 3 tensor
\begin{equation}\label{eq:tensor_L}
  M^{abc}=x^a T^{bc}-x^b T^{ac}.
\end{equation}
However, in the absence of sources (i.e.\ current and charge densities), the stress-energy and angular momentum tensors are divergence-less,
\begin{equation}
  \partial_a T^{ab}=0,\quad \partial_c M^{abc}=0.
\end{equation}
We can define the quantities
\begin{gather}
  P^a=\int_\mathcal{S} T^{ab}\hat{e}_{\,b}^0\, \du^3 r,
  \\
  \label{eq:total_L}
  L^{ab}=\int_\mathcal{S}
  M^{abc}\hat{e}^0_{\,c}\, \du^3 r,
\end{gather}
where $\mathcal{S}$ is a space-like hyper-surface, and $\hat{e}^0$ is the unit normal to $\mathcal{S}$. For the case where $\mathcal{S}$ is a hyper-surface of constant time, this will correspond to the total energy-momentum, and total angular momentum of the fields. Because of the zero divergence, it follows from Gauss' law that these integrals are independent of the choice of hyper-surface (Moller\cite{moller_theory_1952}, pp.~166-168). This is known as Von Laue's theorem\cite{griffiths_resource_2012}, and implies that the totals are time independent.

Next we shall provide a brief oveview of Hertz potentials, as given by Essex\cite{essex_hertz_1977}. Hertz potentials allow one to easily obtain solutions of Maxwell's equations from the scalar wave equation. The formulation is as follows: let $\bm{\Pi}_e,\bm{\Pi}_m$ be two vector fields. The scalar and vector electromagnetic potentials can then be obtained as follows:
\begin{equation}\label{eq:hertz_classical}
  \begin{gathered}
    \phi=-\bm{\nabla}\cdot\bm{\Pi}_e,
    \\
    \bm{A}=\frac{1}{c}\frac{\partial\bm{\Pi}_e}{\partial t}+\bm{\nabla}\times\bm \Pi_m.
  \end{gathered}
\end{equation}
In order to satisfy Maxwell's equations, Hertz vectors satisfy the inhomogeneous wave equations
\begin{equation}\label{}
  \begin{gathered}
    \square \bm{\Pi}_e=4\pi\bm{P},
    \\
    \square \bm{\Pi}_m=4\pi\bm{M}.
  \end{gathered}
\end{equation}
where $\bm{P},\bm{M}$ are the polarisation and magnetisation vectors respectively. In free space, it is thus sufficient that every component of the Hertz vectors satisfies the homogeneous wave equation.

It is known however, that the covariant polarisation and magnetisation are to be expressed as an antisymmetric tensor (Vanderlinde\cite{vanderlinde_classical_2006}, p.~318),
\begin{equation}\label{}
  \mathcal{M}^{ab}=
  \begin{pmatrix}
    0    & P_x  & P_y  & P_z  \\
    -P_x & 0    & M_z  & -M_y \\
    -P_y & -M_z & 0    & M_x  \\
    -P_z & M_y  & -M_x & 0
  \end{pmatrix}.
\end{equation}
This is due to the interchange of electric and magnetic phenomena under Lorentz transformations: as the electric and magnetic fields mix under Lorentz transformations, so must the polarisation and magnetisation. This implies that the covariant formulation of Hertz potentials is given by
\begin{equation}\label{}
  \Pi^{ab}=
  \begin{pmatrix}
    0           & \Pi_{e\,x}  & \Pi_{e\,y}  & \Pi_{e\,z}  \\
    -\Pi_{e\,x} & 0           & \Pi_{m\,z}  & -\Pi_{m\,y} \\
    -\Pi_{e\,y} & -\Pi_{m\,z} & 0           & \Pi_{m\,x}  \\
    -\Pi_{e\,z} & \Pi_{m\,y}  & -\Pi_{m\,x} & 0
  \end{pmatrix}.
\end{equation}
This approach is given in Post\cite{post_general_1967}, p.~68. It is also noted by Cohen and Kegeles\cite{cohen_electromagnetic_1974}, who additionally give the curved space-time generalisation. The Hertz tensor satisfies the inhomogeneous wave equation
\begin{equation}\label{}
  \square \Pi^{ab}=4\pi\mathcal{M}^{ab}.
\end{equation}
The four-potential and fields are then obtained via
\begin{equation}\label{}
  \begin{gathered}
    A^b=\partial_a \Pi^{ab}, \\
    F^{ab}=\partial^a\partial_c \Pi^{cb}-\partial^b\partial_c\Pi^{ca}.
  \end{gathered}
\end{equation}
We note that when expanded into three-vector notation, that this is equivalent to~\eqref{eq:hertz_classical}. Since $\Pi$ is antisymmetric, and mixed partial derivatives commute, we have
\begin{equation}\label{}
  \partial_a \partial_b \Pi^{ab}=0=\partial_a A^a,
\end{equation}
hence the Lorenz gauge condition is automatically satisfied. This simplifies the expression for the field tensor to
\begin{equation}
  F^{ab}=\partial^a\partial_c\Pi^{cb}.
\end{equation}

Lekner\cite{lekner_electromagnetic_2003} employs the magnetic Hertz vector to construct electromagnetic pulses from a single pulse-like wavefunction. We will set $c=1$, and define:
\begin{equation}
  \begin{gathered}
    \rho=\sqrt{x^2+y^2}, \quad z_-=z-t, \\
    r=\sqrt{\rho^2+z^2}, \quad z_+=z+t.
  \end{gathered}
\end{equation}
We will examine two of Lekner's pulses, one of which Lekner finds to be in contradiction of Von Laue's theorem. Both have a wavefunction given by a Ziolkowski type solution\cite{ziolkowski_localized_1989},
\begin{equation}
  \label{eq:psi}
  \psi=\frac{ab}{\rho^2+(a-iz_+)(b+iz_-)}.
\end{equation}
The two pulses are the `CP' and TE pulses, with respective magnetic Hertz potentials given by
\begin{gather}
  \label{eq:cp_a}
  \bm{\Pi}_{CP}=(i,1,0)\psi,
  \\
  \label{eq:te_a}
  \bm{\Pi}_{TE}=(0,0,1)\psi.
\end{gather}
Lekner shows that these pulses can have total momenta less than $c^{-1}$ times their total energies. This implies that a frame can be found where the total momentum is zero. The energy-momentum densities for a general free space solution of Maxwell's equations satisfy (Theory of Reflection\cite{lekner_theory_2016}, p.~493)
\begin{equation}\label{eq:inequality}
  u^2-p^2c^2\geq0.
\end{equation}
The energy density is positive definite, and thus the total energy \textit{must} be strictly greater than zero. However, if the momentum density is completely symmetric, then when integrated over all space it will cancel, leaving a zero net momentum. This can be true even if~\eqref{eq:inequality} is an equality everywhere. Hence, a sufficient condition for a frame with zero total momentum is a frame where the momentum is symmetrically distributed.

Maxwell's equations are still satisfied under the duality transformation (Jackson\cite{jackson_classical_2007}, p.~274)
\begin{align}
  B & \rightarrow E',  \\
  E & \rightarrow -B',
\end{align}
so from given fields $\bm{E}_0,\bm{B}_0$, we are free to construct a new pulse with fields given by
\begin{equation}
  \label{eq:selfdual}
  \begin{gathered}
    \bm{B}=\bm{B}_0-i\bm{E}_0, \\
    \bm{E}=\bm{E}_0+i\bm{B}_0,
  \end{gathered}
\end{equation}
and this will still be a valid solution of Maxwell's equations. Such a pulse is \textit{self-dual}. Since this results in $\bm{E}=i\bm{B}$, if we take the real parts of the fields and calculate the energy and momentum densities, they have the simple expression
\begin{gather}
  \label{eq:u}
  u=\frac{1}{8\pi}(\bm{E}_r^2+\bm{B}_r^2)
  =\frac{1}{8\pi}\bm{B}\cdot\bm{B}^*,
  \\
  \label{eq:s}
  \bm{s}=\frac{1}{4\pi}\bm{E}_r\times\bm{B}_r
  =\frac{1}{4\pi}\bm{B}_r\times\bm{B}_i,
\end{gather}
where the star denotes complex conjugation.

\section{`CP' pulse}\label{sec:cp_pulse}

For the self-dual `CP' pulse, the complex fields are given by
\begin{gather}
  \begin{gathered}
    \label{eq:cp_b_field}
    B_x=\frac{
      4iab\left( {(b+iz_-)}^2
      +\rho^2e^{-2i\phi}\right)
    }{\alpha^3},
    \\
    B_y=\frac{4ab\left( {(b+iz_-)}^2
      -\rho^2e^{2i\phi}\right)
    }{\alpha^3},
    \\
    B_z=\frac{8ab(b+iz_-)\rho e^{-i\phi}}{\alpha^3},
  \end{gathered}
  \\
  \bm{E}=i\bm{B},
  \\
  \alpha\equiv ab+iaz_--ibz_++r^2-t^2.
\end{gather}
For the energy and momentum densities, see Appendix~\ref{app:cp_momentum_densities}.

To find the totals, we integrate the densities over the hyper-surface $t=0$. This can be achieved using spherical coordinates. These integrals have closed-form solutions, and give total energy and momenta of:
\begin{equation}
  \begin{gathered}
    U=\frac{(a+3b)\pi}{8a^2},
    \\
    S_x=S_y=0,
    \\
    S_z=\frac{(a-3b)\pi}{8a^2},
    \\
    U^2-\bm{S}^2=\frac{3b\pi^2}{16a^3}.
  \end{gathered}
\end{equation}
in agreement with Lekner, 2013\cite{lekner_universal_2013}. Note our convention is to use lower case for the densities, and upper case for the totals.

\subsection{Zero Momentum Frame}\label{subsec:zero_momentum_frame}

As we already have explicit expressions for the fields, we shall Lorentz transform via the usual formula (Jackson\cite{jackson_classical_2007}, p.~558)
\begin{equation}
  \label{eq:lorentz_field}
  \begin{gathered}
    \bm{E}'=\gamma(\bm{E}+\bm{\beta}\times\bm{B})
    -\frac{\gamma^2}{\gamma+1}\bm{\beta}
    (\bm{\beta}\cdot\bm{E}), \\
    \bm{B}'=\gamma(\bm{B}-\bm{\beta}\times\bm{E})
    -\frac{\gamma^2}{\gamma+1}\bm{\beta}
    (\bm{\beta}\cdot\bm{B}).
  \end{gathered}
\end{equation}
As we are dealing with fields in space-time we must also substitute the old coordinates as functions of the new coordinates, as usual. We will be boosting along $z$, and as our pulse satisfies $\bm{E}=i\bm{B}$ in all frames, it will suffice to find $\bm{B}$ in the new frame. This is given by
\begin{equation}
  \begin{gathered}
    \bar{B}_x=\gamma(B_x+\beta E_y)=\gamma(B_x+i\beta B_y),
    \\
    \bar{B}_y=\gamma(B_y-\beta E_x)=\gamma(B_y-i\beta B_x),
    \\
    \bar{B}_z=B_z.
  \end{gathered}
\end{equation}

As the total $x$ and $y$ momenta are zero, we can attempt to find a boost along $z$ that will bring the total momentum is zero. Letting an over-bar denote quantities in the zero momentum frame. The boost velocity is found by solving
\begin{equation}
  \bar{S}_z=\gamma(S_z-\beta U)=0,
\end{equation}
which gives us
\begin{equation}
  \beta=\frac{a-3b}{a+3b}.
\end{equation}
Note that as we are constrained via $|\beta|<1$, we must have $|a-3b|<|a+3b|$. Also note that if $a=3b$ the pulse already has zero momentum. The Lorentz transformed fields come out to be:
\begin{gather}
  \begin{gathered}
    \bar{B}_x=\frac{12ia^2b^3(ab-3(\bar{z}_-^2-
    \rho^2e^{-2i\phi})+2i\sigma\bar{z}_-)}{\bar{\alpha}^3},
    \\
    \bar{B}_y=\frac{12a^2b^3(ab-3(\bar{z}_-^2
    +\rho^2e^{-2i\phi})+2i\sigma\bar{z}_-)}{\bar{\alpha}^3},
    \\
    \bar{B}_z=\frac{24a^2b^3\rho e^{-i\phi}(\sigma+3i\bar{z}_-)}{\bar{\alpha}^3},
  \end{gathered}
  \\
  \bar{\bm{E}}=i\bar{\bm{B}},
  \\
  \bar{\alpha}\equiv\sigma(ab(1+\frac{i}{\sigma}(3\bar{z}_-
  -\bar{z}_+))+\bar{r}^2-\bar{t}^2),
  \\
  \sigma\equiv\sqrt{3ab}.
\end{gather}
We can once again find the densities using equations~\eqref{eq:u} and~\eqref{eq:s}, listed in Appendix~\ref{app:cp_momentum_densities}. When integrated, we get totals of
\begin{equation}
  \begin{gathered}
    \bar{U}=\frac{\pi}{4}\sqrt{\frac{3b}{a^3}}=\gamma(U-\beta S_z), \\
    \bar{S}=0, \\
    \bar{U}^2-\bar{\bm{S}}^2=\frac{3b\pi^2}{16a^3}=U^2-\bm{S}^2.
  \end{gathered}
\end{equation}
Thus, we can see that the total energy momentum has transformed as a four-vector, and Von Laue's Theorem holds. Note that this is \textit{not} in agreement with Lekner, 2013\cite{lekner_universal_2013}, which seems to show the pulse violating Von Laue's theorem. We shall discuss this further in Section~\ref{sec:discussion}.

\subsection{Angular momentum}\label{subsec:angular_momentum}

The total angular momentum is found via equations~\eqref{eq:tensor_L} and~\eqref{eq:total_L}. The resulting tensor is antisymmetric, and will have 6 independent components. The space-space components are the total angular momenta,
\begin{equation}
  \label{eq:L_def}
  \bm{L}=\int_S \bm{x}\times \bm{p} \, \du^3 r=\int_{S}\bm{x}\times(\bm{E}\times\bm{B}).
\end{equation}
The time-space components are sometimes referred to as the `boost angular momentum'\cite{cameron_chirality_2017}, and are related to the motion of the center of energy\cite{pryce_mass-centre_1948}.
\begin{equation}
  \label{eq:N_def}
  \bm{N}=\int_S u\bm{x}-ct\bm{p}\, \du^3 r
  =\int_{S}(E^2+B^2)\bm{x}-t\bm{E}\times\bm{B}\,\du^3 r.
\end{equation}
We can see that in the case of a $t=0$ hyper-surface, this reduces to
\begin{equation}
  \bm{N}=\int_{t=0}u\bm{x}\, \du^3 r=\int_{t=0}(E^2+B^2)\bm{x}\,\du^3 r.
\end{equation}
In our case, in the original frame the boost angular momentum is zero, and the total angular momentum is given by
\begin{equation}
  \bm{L}=\frac{\pi b}{4a}\hat{z}.
\end{equation}
As both the total angular momentum and electromagnetic fields are given by antisymmetric rank two tensors, the Lorentz transformation of the mass moment is analogous to that of the electric field, and the angular momentum analogous to that of the magnetic field. As such, we can simply use~\eqref{eq:lorentz_field}, under the substitution
\begin{gather}
  \bm{E}\rightarrow\bm{N}, \\
  \bm{B}\rightarrow\bm{L}.
\end{gather}
This means that since $\bm{L}$ is along $z$, and $\bm{N}$ is zero, the total angular momentum tensor of the pulse is unchanged by boosts along $z$.

\section{TE+iTM pulse}\label{sec:te}

Next, we consider the pulse given by~\eqref{eq:te_a}. From this TE pulse, we shall construct the self-dual TE+iTM pulse via~\eqref{eq:selfdual}. The (complex) fields are then given by
\begin{equation}\label{eq:te_fields}
  \begin{gathered}
    B_x=\frac{4ab\rho \left(e^{-i\phi}(z_++ia)-e^{i\phi}(z_++ib)\right)}{\alpha^3}m
    \\
    B_y=\frac{4ab\rho\left(
    e^{-i\phi}(a-iz_+)+e^{i\phi} (b+iz_-)
    \right)}{\alpha^3}m
    \\
    B_z=\frac{4 a b \left((a-i z_+) (b+i z_-)-\rho ^2\right)}{\alpha^3}m
    \\
    \bm{E}=i\bm{B}m
    \\
    \alpha\equiv ab+iaz_--ibz_++r^2-t^2.
  \end{gathered}
\end{equation}
The expressions for the momentum densities are unwieldy, and shall be omitted, but can easily be obtained via equations~\eqref{eq:u} and~\eqref{eq:s}. When integrated the totals are given by
\begin{equation}\label{}
  \begin{gathered}
    U=\frac{\pi(a+b)}{8ab},
    \\
    S_x=S_y=0,
    \\
    S_z=\frac{\pi(a-b)}{8ab},
    \\
    U^2-\bm{S}^2=\frac{\pi^2}{16ab},
  \end{gathered}
\end{equation}
reproducing Lekner, 2003\cite{lekner_electromagnetic_2003}. The total angular momentum however, is zero, in both the rotational and boost components.

\subsection{Zero momentum frame}
In this case, we can see that if $a=b$ we are already in the zero momentum frame. Otherwise, a boost of
\begin{equation}\label{}
  \bm{\beta}=\frac{a-b}{a+b}\hat{z},
\end{equation}
will put the pulse in its zero momentum frame. Note that the constraint for $|\beta|<1$ gives us the constraint $|a-b|<|a+b|$ for the boost to be valid. Once again, we use~\eqref{eq:lorentz_field} to transform the fields. The resulting expressions are
\begin{equation}\label{}
  \begin{gathered}
    \bar{B}_x=\frac{8\left(y+\frac{1}{\xi}
      (\bar{z}x-i\bar{t}y)\right)}{\bar{\alpha}^3},
    \\
    \bar{B}_y=\frac{8\left(x -\frac{1}{\xi}(\bar{z}y+i \bar{t}x)\right)}{\bar{\alpha}^3},
    \\
    \bar{B}_z=\frac{4\left(\xi-2i\bar{t}-\frac{1}{\xi}(\rho^2-\bar{z}_+\bar{z}_-)\right)}{\bar{\alpha}^3},
    \\
    \bar{\bm{E}}=i\bar{\bm{B}},
    \\
    \bar{\alpha}\equiv\xi-2i\bar{t}
    +\frac{\bar{r}^2-t^2}{\xi},
    \\
    \xi\equiv\sqrt{ab}.
  \end{gathered}
\end{equation}
The expressions for the densities are once again unwieldy, however the values at $t=0$ are given by
\begin{equation}\label{}
  \begin{gathered}
    u=\frac{2 a^2 b^2}{\pi  {\left(a b+\bar{r}\right)}^4},
    \\
    \bm{s}=0.
  \end{gathered}
\end{equation}
When integrated this gives
\begin{equation}\label{}
  \begin{gathered}
    \bar{U}=\frac{\pi}{4\sqrt{ab}},
    \\
    \bar{\bm{S}}=0,
    \\
    \bar{U}^2-\bar{\bm{S}}^2=U^2-\bm{S}^2=\frac{\pi^2}{16ab}.
  \end{gathered}
\end{equation}
This \textit{is} is agreement with Lekner 2013\cite{lekner_universal_2013}. As the momentum density is zero, the rotational angular momentum will also be zero, and as the energy density is spherically symmetric, when we integrate the boost angular momentum, it will be zero,
\begin{equation}\label{}
  \bar{\bm{N}}=\int_{\bar{t}=0}u\bar{\bm{x}}\,\du^3\bar{r}=0.
\end{equation}
Due to Von Laue's theorem, the total momenta and angular momenta are conserved quantities, thus the total angular momenta are zero for all time.

\section{Discussion}\label{sec:discussion}

We have seen that Lekner's result for the Lorentz transformation of the `CP' pulse was in error, this was due to neglecting the bivector nature of the Hertz potentials. To properly Lorentz transform Hertz potentials, one should use~\eqref{eq:lorentz_field}, under the substitution
\begin{equation}\label{}
  \bm{E}\rightarrow\bm{\Pi}_e,\quad \bm{B}\rightarrow\bm{\Pi}_m.
\end{equation}
For both pulses, Lekner transformed the Hertz potentials via\cite{lekner_private}
\begin{equation}\label{}
  \begin{gathered}
    \bm{\Pi}_m(x)\rightarrow\bar{\bm{\Pi}}_m(\bar{x})=\bm{\Pi}_m(x(\bar{x})),
    \\
    \bm{\Pi}_e(x)\rightarrow\bar{\bm{\Pi}}_e(\bar{x})=\bm{\Pi}_e(x(\bar{x})).
  \end{gathered}
\end{equation}
This is correct for the TE+iTM pulse, as in that case $\bm{\Pi}_m$ was parallel to the boost direction, with $\bm{\Pi}_e=0$. This is analogous to how a boost parallel to a magnetic field, with zero electric field, will not change the fields. However, for the `CP' pulse, this is incorrect. $\bm{\Pi}_m$ is perpendicular to the boost; thus not only will $\bm{\Pi}_m$ be strengthened by a factor of $\gamma$, but $\bm{\Pi}_e$ will no longer be zero.

\begin{acknowledgments}
  The author would like to thank Prof.\ John Lekner for his feedback and constructive criticism.
\end{acknowledgments}

\appendix
\section{`CP' momentum densities}\label{app:cp_momentum_densities}
For the `CP' pulse in the original frame, the four momentum densities are given by
\begin{equation}
  \begin{gathered}
    u=\frac{4a^2b^2{\left(b^2+\rho^2+z_-^2\right)}^2
    }{\pi|\alpha|^6},
    \\
    s_x=-\frac{8a^2b^2{\left(b^2+\rho^2+z_-^2\right)}
    (by+z_-x)}{\pi|\alpha|^6},
    \\
    s_y=\frac{8a^2b^2{\left(b^2+\rho^2+z_-^2\right)}
    (bx-z_-y)}{\pi|\alpha|^6},
    \\
    s_z=-\frac{4a^2b^2(b^2-\rho^2+z_-^2)
      (b^2+\rho^2+z_-^2)}{\pi|\alpha|^6}.
    \\
    \alpha\equiv ab+iaz_--ibz_++r^2-t^2.
  \end{gathered}
\end{equation}

In the zero momentum frame, the densities are given by:
\begin{equation}
  \begin{gathered}
    \bar{u}=\frac{36a^4b^6{(ab+3(\bar{z}_-^2+\rho^2))}^2
    }{\pi|\bar{\alpha}|^6},
    \\
    \bar{s}_x=-\frac{72a^4b^6(ab+3(\bar{z}_-^2+\rho^2))
      (3\bar{z}_-x+\sigma y)}{\pi|\bar{\alpha}|^6},
    \\
    \bar{s}_y=\frac{72a^4b^6(ab+3(\bar{z}_-^2+\rho^2))
      (\sigma x-3\bar{z}_- y)}{\pi|\bar{\alpha}|^6},
    \\
    \bar{s}_z=-\frac{36a^4b^6(ab+3(\bar{z}_-^2+\rho^2))
      (ab+3(\bar{z}_-^2-\rho^2))}{\pi|\bar{\alpha}|^6}.
    \\
    \bar{\alpha}=\sigma(ab(1+\frac{i}{\sigma}(3\bar{z}_-
    -\bar{z}_+))+\bar{r}^2-\bar{t}^2).
    \\
    \sigma=\sqrt{3ab}.
  \end{gathered}
\end{equation}

The rotational and boost angular momentum densities can easily be found from the above expressions and equations~\eqref{eq:L_def} and~\eqref{eq:N_def}.

\bibliography{zotero,custom}

\end{document}